# Spin Diode based on Fe/MgO Double Tunnel Junction

A. Iovan[1], S. Andersson[1], Yu. G. Naidyuk[1], A. Vedyaev[2], B. Dieny[3], and V. Korenivski[1]

[1]*Nanostructure Physics, Royal Institute of Technology, 10691 Stockholm, Sweden*
[2]*Lomonosov State University, Department of Physics, Moscow 119699, Russia*
[3]*SPINTEC, URA CEA/CNRS, Grenoble, France*

**We demonstrate a spin diode consisting of a semiconductor free nano-scale Fe/MgO-based double tunnel junction. The device exhibits a near perfect spin-valve effect combined with a strong diode effect. The mechanism consistent with our data is resonant tunneling through discrete states in the middle ferromagnetic layer sandwiched by tunnel barriers of different spin-dependent transparency. The observed magneto-resistance is record high, ~4000%, essentially making the structure an on/off spin-switch. This, combined with the strong diode effect, ~100, offers a new device that should be promising for such technologies as magnetic random access memory and re-programmable logic.**

Recently discovered spin-valves and magnetic tunnel junctions are finding numerous applications as sensors and memory elements[1-5]. Giant magneto-resistance obtained in these devices motivates the great current research effort on developing spin-based diodes and transistors[6]. Implementing these magnetic logic elements in semiconductors is challenging, however, due to such fundamental issues as efficient spin injection and ferromagnetism at room temperature[7-10]. Double tunnel junctions (DTJ's) having the two tunnel barriers of different transparency can exhibit highly asymmetric conduction for different polarity bias, i.e., act as a current rectifier or a diode[11]. Such current rectification has been observed in double tunnel junctions[12,13] for the case of *spin-independent* conductivity (i.e., vanishing magneto-resistance, MR≈0). Current rectification and MR have also been observed for *single* asymmetric tunnel barriers[14], however with limited rectification ratios (RR~10). We have recently demonstrated a very strong diode effect (RR~$10^4$) in metal/oxide DTJ's[15]. In this report we demonstrate a record high tunnelling MR≈4000% combined with a high RR~100 in the same DTJ, making the device an efficient hybrid of a spin switch and a diode. The material system we use in this demonstration is MgO-based magnetic tunnel junctions, where MR values 100-600% have been reported recently[16,17,18,19].

F1/I1/F2/I2/N samples of structure Si/SiO/Fe(50nm)/MgO(3nm)/Fe(2nm)/MgO(2nm)/Au(30nm) were deposited using dc magnetron sputtering at the base pressure of 1.3e-8 mbar. The argon pressure of 4.0e-3 mbar was used for deposition of the Fe layers. The magnesium oxide was reactively sputtered by adding 4.0e-4 mbar of oxygen to the Ar sputter gas during the sputter deposition of Mg. The different thickness of the MgO layers is responsible for the asymmetry in the transparency of the two tunnel barriers sandwiching the middle Fe layer. The nanometer thin middle Fe insulated by MgO from the outer electrodes is designed to have discrete electron states, with the level spacing of the order of 100 meV. The stacks were patterned into nanopillars using a 150 nm ZEP520A positive resist. The resist was spun on to the multilayer samples and hardened by baking. A rectangular matrix of lines was drawn by e-beam lithography in the resist, which was subsequently developed to serve as a hard mask. This hard mask was then transferred on to the TJ stacks using Ar ion beam etching. The nanopillar fabrication process was finalized by removing the resist mask with oxygen plasma. The resulting structure is a large array of DTJ vertical stacks, separated by trenches etched through the top Au electrode, the top MgO barrier, and the middle Fe electrode, down to the bottom MgO layer. Since the etching rate of MgO is much lower than that of metals, we could reliably stop the ion etch at the thicker bottom MgO layer using proper etch rate calibrations. Thus, the patterned stack is capped with Au, which acts as the top electrode of the DTJ and defines the lateral size of the junction. The lateral and vertical profile of the samples was characterized using a scanning tunneling microscope (STM), as shown in Fig. 1a. More details on the fabrication and STM characterization methods used in this study can be found elsewhere[20]. The thick Fe layer acts as the common bottom electrode for the DTJ array as a whole. The main advantage of this patterning process, with the etch stopped at the bottom MgO barrier, is that it minimizes shortcuts at the TJ perimeter due to possible re-deposition compared to the case where the whole stack is patterned. By controlling the in-plane geometry of the nanopillars we control the shape anisotropy of the middle Fe layer, which allows to reliably separate the switching fields of the magnetically softer continuous bottom Fe layer and the magnetically harder patterned middle Fe layer.

The samples were measured using the point contact technique[21,22,23] directly in liquid helium. This technique allows spectroscopic transport measurements by establishing stable mechanical contacts of typical size ~10 nm, much smaller than the nanopillar size. The resistance of such point contacts is typically 1 to 10 Ω, negligible on the scale of the TJ resistance. The advantage of using such nano-mechanical technique is that possible damage to the sensitive regions of the DTJ from fabrication of the top contact is avoided, and a very large number of stacks can be screened for a given sample in the same cool down. The stability of the contacts that were subsequently analyzed for quantum transport was verified by repeating the current-voltage and conductance-field sweeps several times for a given contact/stack.

The direct evidence of a DTJ behavior, with the current flowing through discrete quantum well states in the center Fe electrode, is the measured quantized conduction shown in Fig. 2. The steps in current versus bias voltage and the associated peaks in the differential conductance are spaced by ~100 mV, which is consistent with the energy spacing expected for the middle Fe layer of ~2 nm in thickness[24]. Since the probability of tunneling decreases exponentially with the thickness of the tunnel barrier, the two tunnel barriers of thickness 3 and 2 nm have substantially different transparency. In this case theory predicts[11] asymmetric conduction for different polarity bias. We indeed observe pronounced transport asymmetry, with the rectification ratio RR=$|I(+V_b)/I(-V_b)|$ of up to 100, as shown in Fig. 3a. This demonstrates a strong diode effect for a magnetic double tunnel junction.

All samples measured showed a dependence of the resistance on the applied magnetic field, which is a good indication of the quality of the bottom MgO barrier and the patterning processes as a whole. The magneto-resistance at a positive bias of 40 mV shown in Fig. 3c exhibits the behavior typical of spin-valves: the conductance is high at high fields where the magnetizations of the Fe layers are parallel, and low at intermediate fields where the magnetizations are antiparallel. Interestingly, reversing the bias direction results in an inverted magneto-resistance, as shown in Fig. 3b for $V_b$=-50 mV. The switching fields of ~4 mT and

~40 mT are the same as those for the positive bias configuration. The lower switching field corresponds to the reversal of the magnetization of the un-patterned bottom Fe layer. The higher switching field corresponds to the reversal of the magnetization in the middle Fe electrode and is consistent in magnitude with the shape anisotropy field due to the in-plane patterning (170x800 nm$^2$ in this case). The fact that the switching of the middle electrode is sharp is an additional evidence that the thin Fe layer is continuous, since nano-granular Fe would saturate in much higher fields (~1 T for spherical particles) over a much broader field range (0 to 1 T). Thus, for negative bias the conductivity is low at high fields where the magnetizations are parallel and high at intermediate fields where the magnetizations are antiparallel. The associated MR is negative and approaches 4000%, as shown in Fig. 3b.

Our test samples with single tunnel barriers universally show a positive MR as expected for spin-valves[25,26]. Thus, the inverted MR is the signature of the magnetic DTJ, which is in the quantum transport regime (see Fig. 2). This behaviour can originate from the spin dependence of the quantized energy states in the middle electrode[27]. It has been predicted that such discrete states can shift as a function of the magnetic misalignment of the two ferromagnetic electrodes[28,29,30]. Our interpretation of the observed behaviour is as follows.

As shown in Fig. 2, the current through the DTJ as a function of voltage is quantized in the form of a staircase. This behaviour is expected for electron transmission through discrete quantum well states (QWS's), with the current in the general case given by

$$J_\uparrow = \frac{2e^2}{\pi\hbar} \int_{\mu_0}^{\mu_0+eV} dE \sum_n \frac{\Gamma^n_{L\uparrow} \cdot \Gamma^n_{R\uparrow(\downarrow)}}{(E_n - E)^2 + \left(\Gamma^n_{L\uparrow} + \Gamma^n_{R\uparrow(\downarrow)}\right)^2} \cdot \quad (1)$$

Here $J_\uparrow$ is the spin-up current component, with the electrons transmitted through QWS's of energy $E_n$ and width $\Gamma^n$, $V$ is the bias voltage applied to the DTJ, and $\mu_0$ is the chemical potential in the outer electrodes at zero bias. Indeed, at sufficiently low temperature the resonant terms in Eq. 1 integrated over all available electron energies produce a step-like I(V) characteristics IVC, of the kind we observe experimentally.

To explain the observed magneto-resistance (Fig. 3) it will be sufficient to consider the qualitative energy diagram shown in Fig. 4. Thus, Fig. 4a depicts the antiparallel configuration, with the electrons flowing rightward from Fe to Au, and the bias voltage taken to match one of the QW states. The electron transmission is resonant, resulting in a high conductance in this case. Upon reversal of one of the magnetic electrodes into the parallel magnetic state of the left Fe/MgO/Fe junction, this junction resistance decreases as does the voltage drop across it. With the fixed bias voltage across the double junction this leads to a shift of the energy levels in the middle electrode, as illustrated in Fig. 4b, such that the transmission becomes off-resonant and the conductance of the DTJ as a whole decreases. This consideration explains the negative MR, which is the opposite of the normal spin-valve effect where the high conductance corresponds to the parallel magnetic configuration.

Reversing the polarity of the bias shifts the QW levels in the middle electrode such that the electron transmission is off-resonant for both magnetic states of the left junction, as illustrated in Fig. 4c,d. Since the left MgO barrier is thicker, it is the left magnetic junction that contributes most to the total resistance of the DTJ. One therefore expects for this bias configuration a TMR of type normal for Fe/MgO single junctions, which is indeed observed - the conduction is high for the parallel magnetic state and low for the antiparallel state.

As expected for a non-resonant transmission, the magnitude of the measured positive MR corresponds well with the values reported for Fe/MgO/Fe single junctions of 100-800%. The observed negative, resonant MR on the other hand is an order of magnitude higher. This substantial difference in magnitude in addition to the difference in sign of the observed MR provides an important additional confirmation of our interpretation of the observed magneto-conductance effect as due to resonant spin dependent tunnelling through quantum well states of the F/F/N double tunnel junction.

In conclusion, the measured magneto-resistance is record high, essentially making the structure an on/off spin-switch. This, combined with the strong diode effect, offers a new device that should be promising for such technologies as MRAM and re-programmable logic.

**References:**

[1] M. N. Baibich, J. M. Broto, A. Fert, F. Van Dau Nguyen, F. Petroff, P. Etienne, G. Creuzet, A. Friederich and J. Chazelas, *Phys. Rev. Lett.* **61** 2472 (1988).

[2] B. Dieny, V. S. Speriosu, S. S. P. Parkin, B. A. Gurney, D. R. Wilhoit, and D. Mauri, *Phys. Rev. B* **43**, 1297 (1991).

[3] J. S. Moodera, L. R. Kinder, T. M. Wong and R. Meservey, *Phys. Rev. Lett.* **74** 3273 (1995).

[4] T. Miyazaki and N. Tezuka, *J. Magn. Magn. Mater.* **139**, L231 (1995).

[5] S.P.P. Parkin *et al*, *J. Appl. Phys.* **85,** 5828 (1999).

[6] I. Zutic, J. Fabian, S. Das Sarma, *Rev. Mod. Phys.*, **76**, 323 (2004).

[7] G. Schmidt, D. Ferrand, L. W. Molenkamp, A. T. Filip, and B. J. van Wees, *Phys. Rev. B* **62**, R4790 (2000).

[8] H. Ohno, *Science* **281,** 951 (1998).

[9] H. Ohno, D. Chiba, F. Matsukura, T. Omiya, E. Abe, T. Dietl, Y. Ohno and K. Ohtani *Nature* **408,** 944 (2000).

[10] Y. D. Park, A. T. Hanbicki, S. C. Erwin, C. S Hellberg, M. Sullivan, J. E. Mattson, T. F. Ambrose, Wilson, G. Spanos and B. T. Jonker, *Science* **295,** 651 (2002).

[11] M. Chshiev, D. Stoeffler, A. Vedyayev, and K. Ounadjela, Europhys. Lett. **58**, 257-263 (2002).

[12] C. Tiusan, M. Chshiev, A. Iovan, V. da Costa, D. Stoeffler, T. Dimopoulos, and K. Ounadjela, *Appl. Phys. Lett*. **79**, 4231 (2001).

[13] A. Iovan, *PhD thesis*, IPCMS, Strasbourg (September 2004).

[14] C. de Buttet, M. Hehn, F. Montaigne, C. Tiusan, G. Malinowski, A. Schuhl, E. Snoeck, and S. Zoll, *Phys. Rev. B* **73**, 104439 (2006).

[15] A. Iovan, D. B. Haviland and V. Korenivski, *Appl. Phys. Lett*. **88**, 163503 (2006).

[16] W. H. Butler, X.-G. Zhang, T. C. Schulthess, and J. M. MacLaren, *Phys. Rev. B* **63**, 054416 (2001).

[17] S. S.P. Parkin, C. Kaiser, A. Panchula, P. M. Rice, B. Hughes, M. Samant, S.H. Yang, *Nature materials* **3**, 862 (2004).

[18] S. Yuasa, T. Nagahama, A. Fukushima, Y. Suzuki, K. Ando, *Nature Materials* **3**, 868 (2004).

[19] S. Yuasa, A. Fukushima, H. Kubota, Y. Suzuki, K. Ando, *Appl. Phys. Lett.* **89**, 042505, (2006).

[20]A. Iovan, K. Lam, S. Andersson, S. S. Cherepov, D. B. Haviland, V. Korenivski, *IEEE Trans. Magn*. **43**, 2818 (2007).

[21]Yu. G. Naidyuk and I. K. Yanson, *Point-Contact Spectroscopy*, Springer Series in Solid-State Sciences, Vol. 145 (Springer Science+Business Media, Inc, 2005).

[22]I. K. Yanson, Yu.G. Naidyuk, D. L. Bashlakov, V.V. Fisun, O. P. Balkashin, V. Korenivski, A. Konovalenko, and R. I. Shekhter, *Phys. Rev. Lett*. **95,** 186602 (2005).

[23]I. K. Yanson, Yu. G. Naidyuk, V. V. Fisun, A. Konovalenko, O. P. Balkashin, L. Yu. Triputen, and V. Korenivski. *Nano Lett.* **7**, 927 (2007).

[24]T. Nozaki, N. Tezuka, and K. Inomata, *Phys. Rev. Lett*. **96**, 027208 (2006).

[25] M. Julliere: *Phys. Lett* A**, 54** 225 (1975).

[26] J. C. Slonczewski, *Phys. Rev. B* **39**, 6995 (1989).

[27] K. Yakushiji, F. Ernult, H. Imamura, K. Yamane, S. Mitani, K. Takanashi, S. Takahashi, S. Maekawa, and H. Fujimori, *Nat. Mater*. **4**, 57 (2005).

[28] J. Barnas and A. Fert, *Phys. Rev. Lett.* **80**, 1058 (1998).

[29] A. Brataas, Y. V. Nazarov, J. Inoue, and G. E. W. Bauer, *Phys. Rev. B* **59**, 93 (1999).

[30] J. Barnas, and A. Fert, *Europhys. Lett.* **44**, 85−90 (1998).

**Figure captions:**

Fig. 1. (a) STM image of an array of nano-pillar junctions and schematic of the point-contact measurement technique. (b) Illustration of a magnetic double tunnel junction with a resonant transmission state in the middle quantum well.

Fig. 2. Current ($I$) and conductance ($dI/dV$) as a function of bias voltage ($V$) for a typical DTJ in the quantum transport regime. The numbers in the caption give the thickness of the individual layers of the DTJ in nanometers.

Fig. 3. (a) Asymmetric current voltage characteristics of a double tunnel junction in anti-parallel magnetic state, exhibiting a strong diode effect. Magnetoresistance for a negative bias of -50 mV (b), and a positive bias of 40 mV (c). The inset to (a) shows a proposed circuit symbol for the device.

Fig. 4. Spin-dependent energy diagram of the DTJ: (a) negative bias, anti-parallel magnetic state; (b) negative bias, parallel magnetic state; (c) positive bias, anti-parallel state; (d) positive bias, parallel magnetic state. The bias voltage across the left spin-dependent junction ($V_L$) is a function the bias polarity, DTJ asymmetry and the magnetic state of the left junction, and is generally different for the four configurations of (a)-(c). Only configuration (a) corresponds to resonant transmission through one of the discrete states in the middle electrode (QWS1).

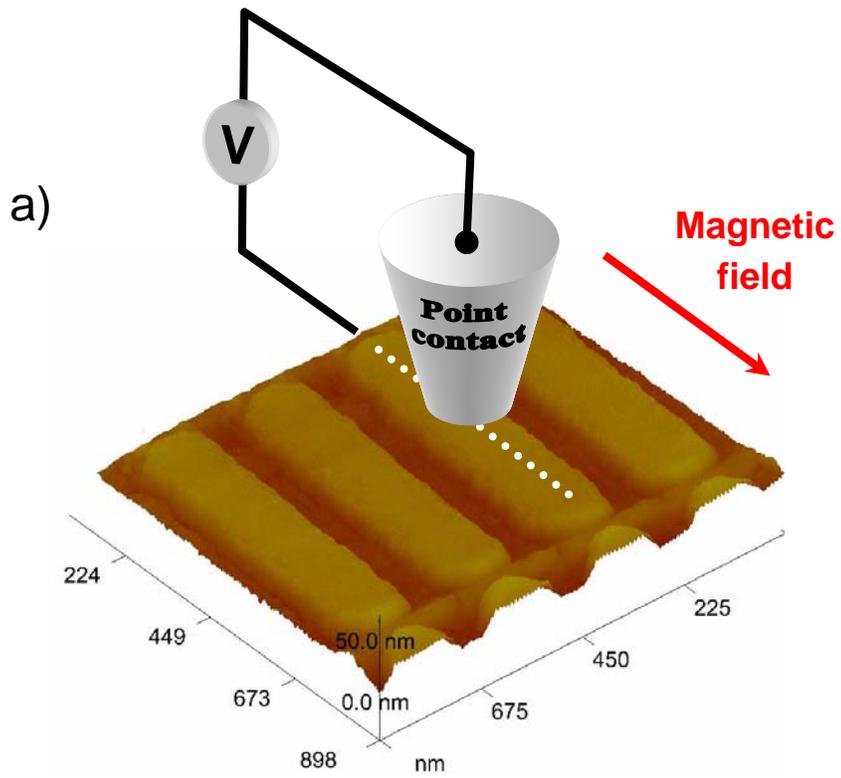

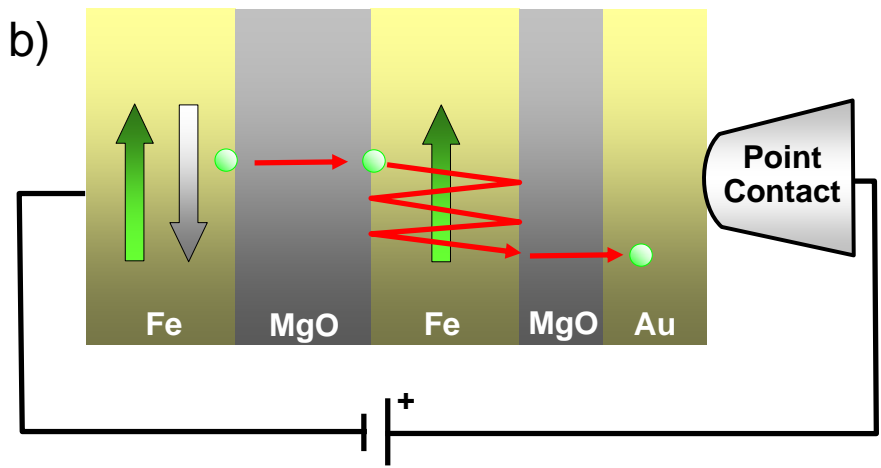

Fig. 1

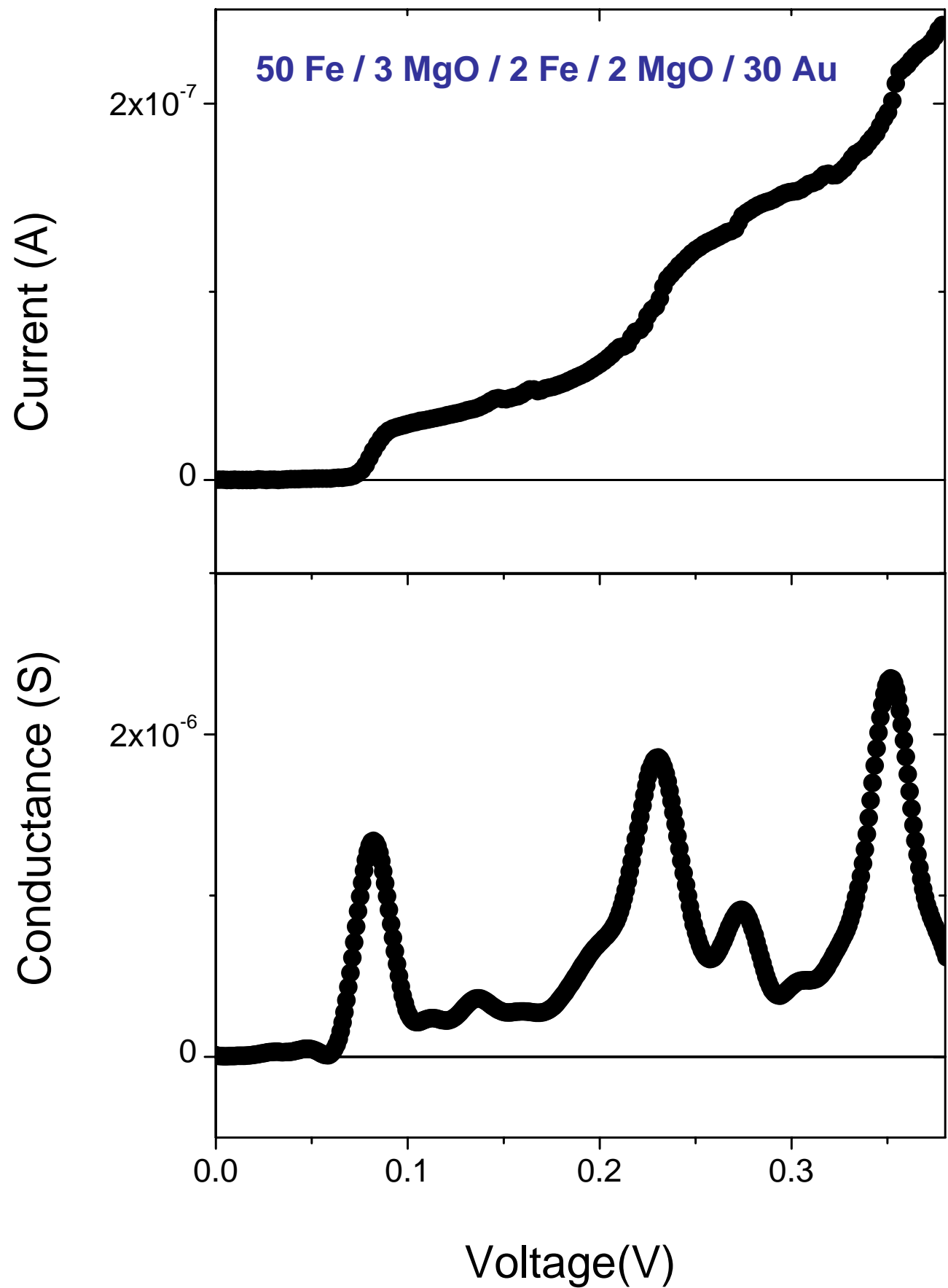

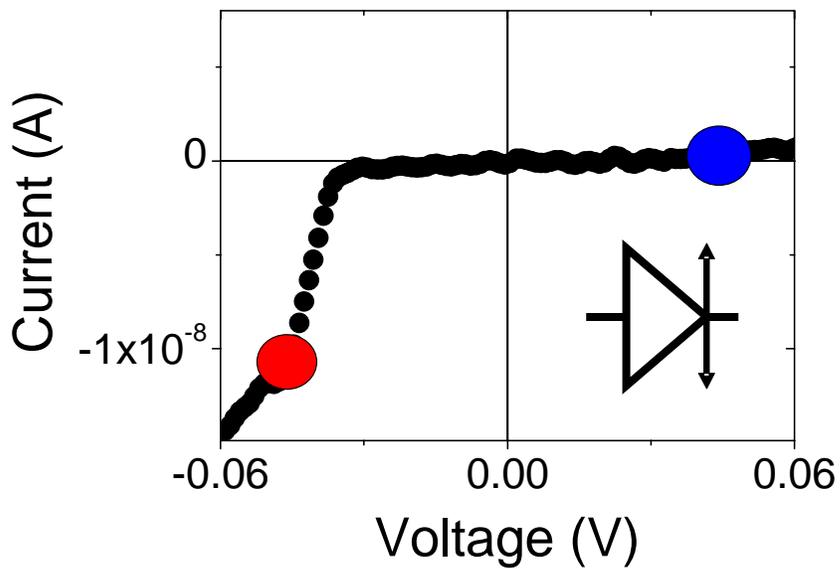
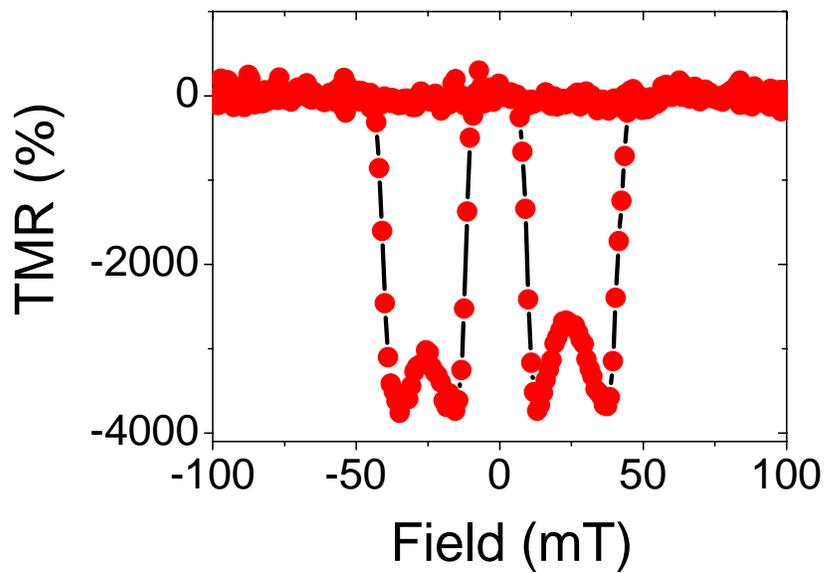
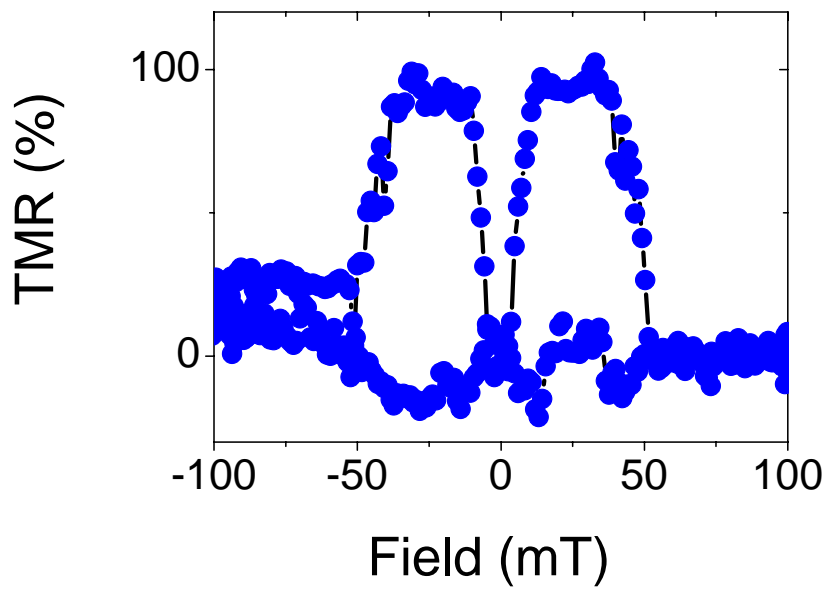

Fig. 3

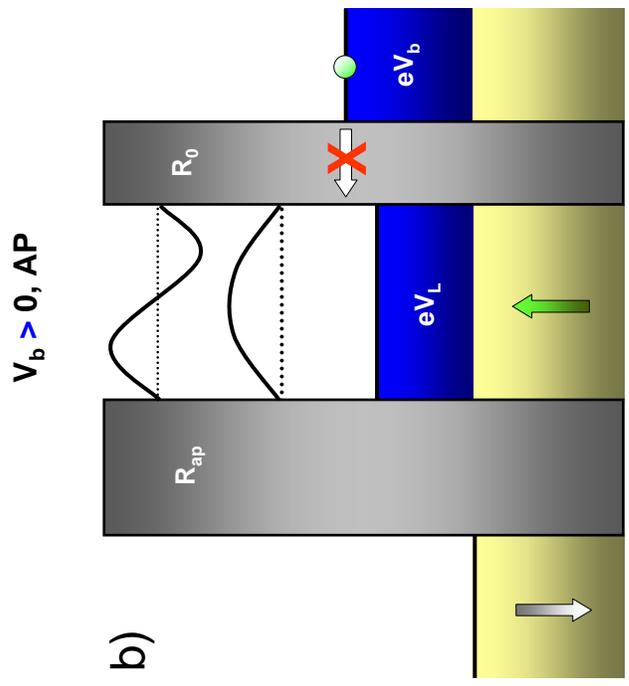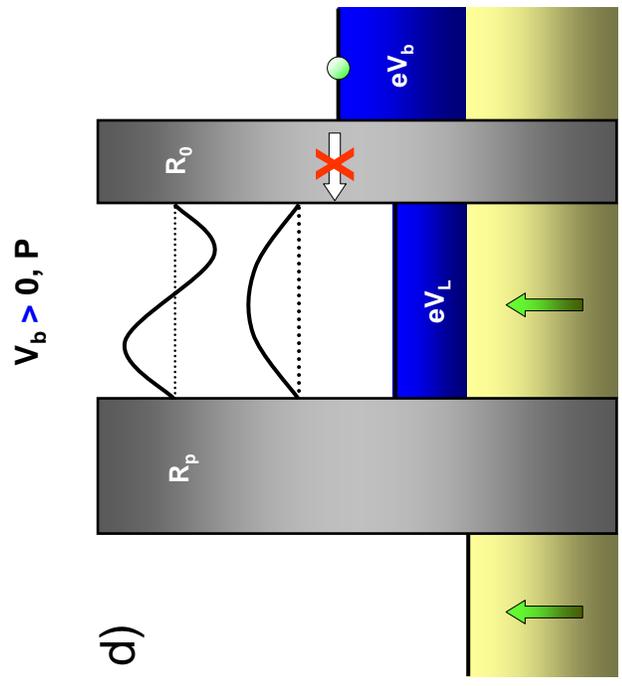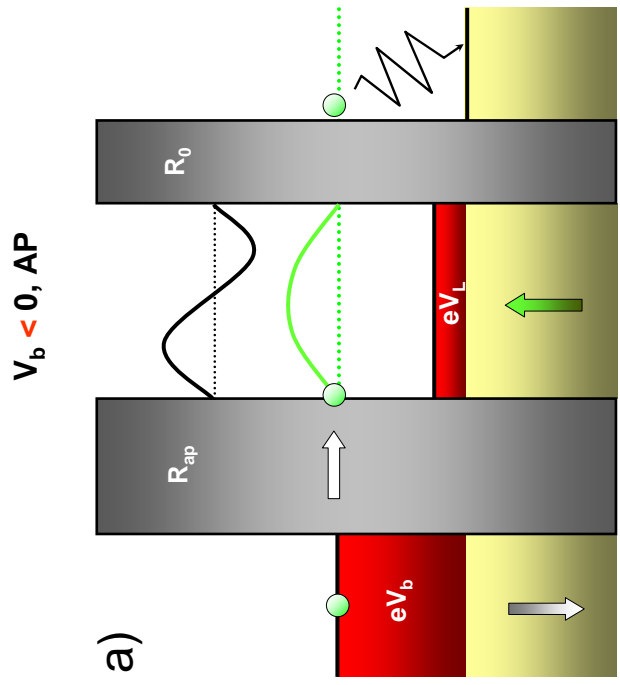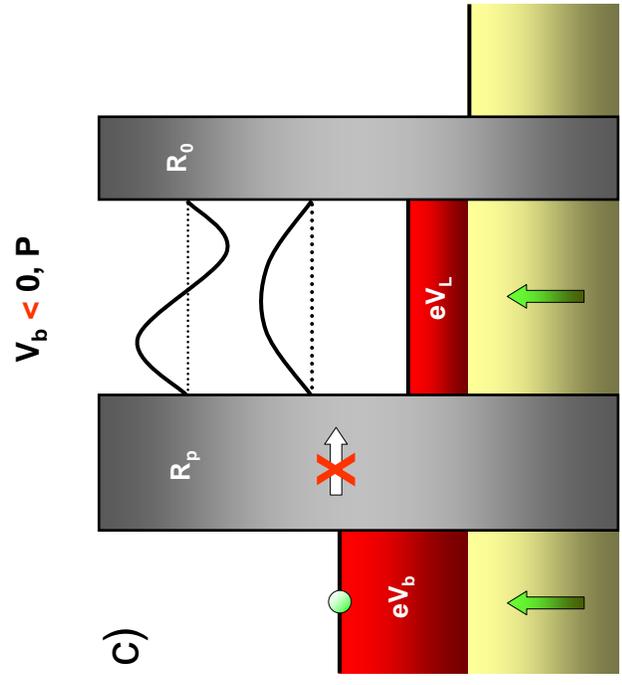

Fig. 4